\documentclass[a4paper]{article}

\usepackage{epsfig}
\usepackage{lscape}

\begin{document}

\title{Basic characterization of highly enriched uranium by gamma spectrometry}

\author{%
Cong Tam Nguyen\thanks{E-mail: tam@iki.kfki.hu}
\ and J\'ozsef Zsigrai\thanks{E-mail: zsigrai@sunserv.kfki.hu}\\
{\small \it Institute of Isotopes of the Hungarian Academy of Sciences}\\
{\small \it H-1525 Budapest, P.O. Box 77, Hungary} }

\maketitle

\begin{abstract}
Gamma-spectrometric methods suitable for the characterization of highly enriched uranium samples encountered in illicit trafficking of nuclear materials are presented. In particular, procedures for determining the $^{234}$U, $^{235}$U, $^{238}$U, $^{232}$U and $^{236}$U contents and the age of highly enriched uranium are described. Consequently, the total uranium content and isotopic composition can be calculated. For determining the $^{238}$U and $^{232}$U contents a low background chamber was used. In addition, age dating of uranium was also performed using low-background spectrometry.
\medskip

Keywords: Nuclear forensics, Low background gamma-spectrometry, Illicit trafficking of nuclear materials, Fissile material cut-off treaty

\medskip
PACS:  29.30.Kv; 28.60.+s; 29.90.+r

\end{abstract}

\section{Introduction}

The detection of illicit trafficking of nuclear materials, including ``nuclear smuggling'' is necessarily followed by forensic investigations in order to determine the possible origin of the material. These investigations involve, beside classical forensic analysis, also the techniques of nuclear forensic science. Nuclear forensic investigations are carried out by laboratories capable of hand\-ling and analyzing nuclear material of unknown origin, or ``nuclear forensic laboratories''.

At least three levels of nuclear forensic capabilities can be distinguished \cite{INFLCharter,MAP}. These are: categorization, characterization, and full attribution of materials to the point of loss-of-control. 

Categorization means the determination of the type of the investigated material (``low-enriched'' uranium, ``high-enriched'' uranium, plutonium, irradiated materials, other radioactive substances), in order that a threat assessment can be done. Categorization is usually performed by high-resolution gamma-spectroscopy, possibly using portable equipment so that a fast threat assessment can be done on the spot where illicit trafficking is detected. 

Characterization involves a more precise and more detailed determination of isotopic composition, crystal structure, trace-element analysis, age dating, particle size analysis etc. Some of the methods used for characterization of nuclear material of unknown origin are mass spectroscopy, alpha spectroscopy, X-ray diffraction analysis, electron microscopy, etc. 

Upon characterizing nuclear material of unknown origin, some questions related to the production processes and intended end use may be addressed. If an appropriate knowledge base is also available, full attribution of materials to the point of loss-of-control might be possible.

In this work we wish to show that simple gamma spectroscopy can be used not merely for categorization, but also for basic characterization of highly enriched uranium\footnote{In this work the term ``enriched uranium'' is used to denote uranium with a relative isotope abundance in $^{235}$U larger than the natural $^{235}$U content of uranium. The expression ``highly enriched uranium'' (HEU) has various meanings in the literature, but here we will only use it to denote isotopically enriched uranium with more than 80\% $^{235}$U, contrary to most of its uses elsewhere.} (HEU) of unknown origin. In particular, it is shown how gamma spectroscopy can be used for very accurate determination of the age and the isotopic composition of HEU. These are properties which are usually assessed by mass spectroscopy, which is relatively complicated, requires sophisticated (and expensive) equipment and special sample preparation is needed. On the contrary, gamma spectroscopy is relatively simple and, since it is non-destructive, it does not require any special sample preparation. Therefore, although gamma-spectroscopy is not capable of all types of analysis done by mass spectroscopy (e.g. trace element analysis), in nuclear forensics it might be the preferred choice for laboratories that do not possess an appropriate mass spectrometer or do not need the additional information supplied by mass spectrometry.

In 2001 the International Technical Working Group for Combating Illicit Trafficking of Nuclear Materials (ITWG) organized an inter-laboratory comparison exercise (Round Robin Exercise - RRE) in order to assess and compare means and methods of forensic analysis on samples of seized nuclear material \cite{Dudderetal,Dudderetal-Karlsruhe}. In the Exercise samples containing about 2 g of a highly enriched uranium oxide powder were sent to the participating laboratories. The participants were asked to determine the properties of this material, relevant for its forensic analysis. The results were expected to be returned to the organizers in periods of 24 hours, 1 week and 2 months.

In this paper the gamma-spectrometric methods which can be used to determine the $^{234}$U, $^{235}$U, $^{238}$U, $^{232}$U, and $^{236}$U, as well as the $^{232}$Th and $^{241}$Am contents and the age of HEU are reviewed. In order to determine the age of the sample, the activity of $^{214}$Bi was measured. The applicability of the methods is demonstrated by determining these quantities  for the sample received within the Round Robin Exercise.

Three high-purity germanium (HPGe) detectors were used for the measurements.
The $^{234}$U and $^{235}$U content of the sample were determined using a large-area planar high-purity germanium  detector. A procedure for self-absorption correction was applied and the count rates per unit mass of the relevant gamma lines of $^{234}$U and $^{235}$U were determined. The isotope contents were then calculated by comparing these count rates to the corresponding ones obtained from the measurement of a reference sample.

A coaxial HPGe detector in a low background iron chamber was used to measure the activities of $^{232}$U, $^{238}$U, $^{232}$Th and $^{214}$Bi, based on the knowledge of the absolute efficiency curve of the detector in the relevant energy interval. 

Finally, the activities of $^{236}$U and $^{241}$Am were estimated from the spectra taken by a medium-area planar HPGe detector, using an intrinsic efficiency calibration method. 

From the above measurements the activity ratio $^{214}$Bi/$^{234}$U was calculated, and using this result the age of the sample was also determined. The presence of $^{232}$U and $^{236}$U indicates that the investigated material was produced by reprocessing spent nuclear fuel.

At the end of the paper a discussion of the results is given, together with a comparison of the measured values to those obtained using mass-spectrometry by other laboratories participating in the Round-Robin exercise.

\section{Determination of the $^{234}$U and $^{235}$U content}
\label{U234}

The $^{234}$U and $^{235}$U contents of the Round Robin Material (``RRM'') were determined using a 20 cm$^2$ planar HPGe detector (Canberra GL2020 with active diameter of 50.5 mm and thickness of 20 mm). A reference material (``RFM'') was used for calibration. The RFM was U$_3$O$_8$ powder of known isotopic composition ($^{235}$U: 90.6 $\pm$ 1.5\%; $^{238}$U: 8.35 $\pm$  0.20\% and $^{234}$U: 1.02 $\pm$  0.07\%). This means that in 1 g of the RFM there is a total of 0.846 g of uranium consisting of $0.766\pm0.12$ g $^{235}$U, $0.070\pm0.001$ g $^{238}$U and $0.0086\pm 0.0004$ g of $^{234}$U. The reference material was transported to Hungary before 1960, so its age at the time of the measurement (June 2001) was more than 40 years. Both nuclear materials (the RRM and the RFM) were placed separately into a cylindrical polyethylene container of 2.9 cm inner diameter.

\begin{figure}[htbp]
\begin{center}
\epsfig{figure=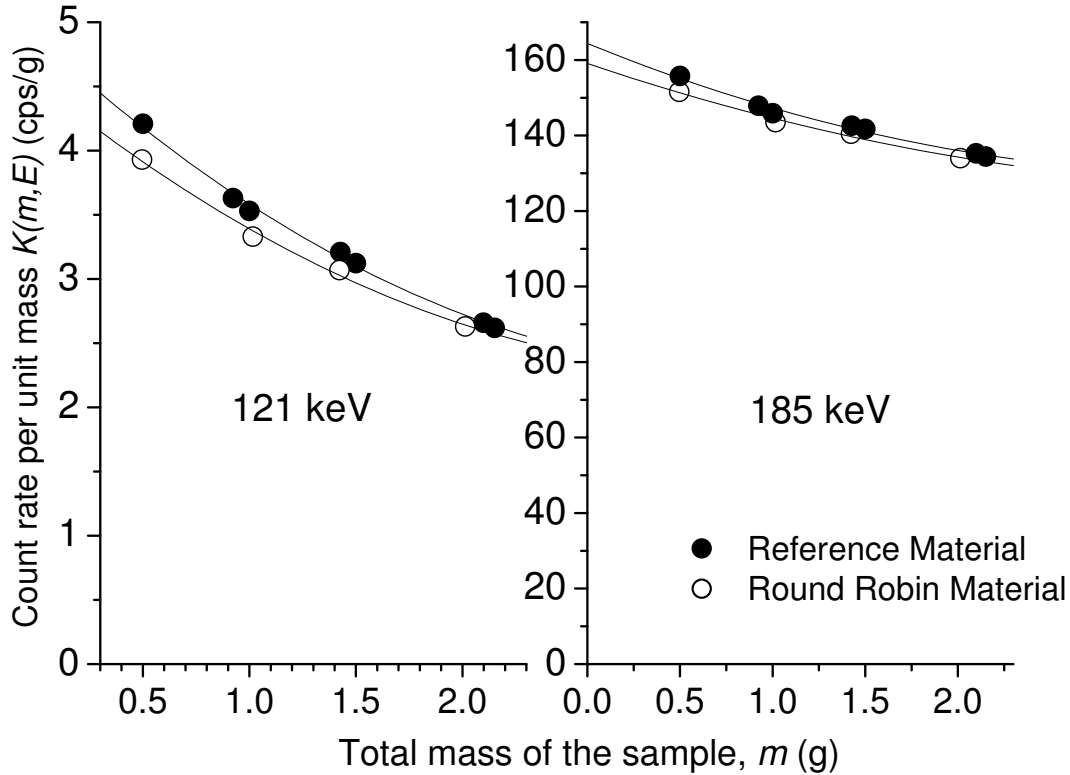, width=\textwidth}
\caption{Count rate of 120.9 and 185.7 keV per unit mass of the sample as a function of the total mass of the sample.}
\label{kev121and185}
\end{center}
\end{figure}

The spectra of the samples were measured at a 10.8 cm distance from the detector. Four measurements with amounts of approximately 0.5, 1, 1.5 and 2 g were carried out with each material. The count rates of the gamma-transitions at 120.9 keV ($^{234}$U) and 185.7 keV ($^{235}$U) were standardized in unit mass as the ratio, $K(m,E)$, of the count rate to the total used mass of the sample, $m$, and then they were plotted versus $m$ (see Fig. \ref{kev121and185}).

The count rates per unit mass at energy $E$, $K(m,E)$, were fitted with the function

\begin{equation}
K(m,E)=K_0(E){{1-e^{-a_Em}}\over{a_Em}}\ , \label{Km}
\end{equation}
which models the law of self-absorption within the sample. The parameter $K_0(E)$ is the value of the curve at $m=0$ and physically it represents the net count rate at energy $E$ of 1 g sample, corrected for self-absorption. The parameter $a_E$ provides information about the matrix of the sample. The $^{234}$U and $^{235}$U contents of 1 g of the sample were derived by comparing the count rate per unit mass of the RRM, $K_0(E)$, to that of the RFM, $K'_0(E)$. More precisely, the $^{234}$U content of the Round-Robin material, $M(U234)$, was calculated by inserting the corresponding data into the formula 
\begin{equation}
M(U234)=M'(U234){{K_0(120.9keV)}\over{K'_0(120.9keV)}}\ , \label{M}
\end{equation}
where the primed quantities correspond to the reference material. An analogous formula was used for the the $^{235}$U content. A non-linear least squares fit was performed to obtain the parameters $K_0(120.9keV)$, $K'_0(120.9keV)$, $K_0(185.7keV)$, and $K'_0(182.7keV)$. In this way the contents of $^{234}$U and $^{235}$U in 1 g of the RRM were found to be 0.00807$\pm$0005 g/g and 0.754$\pm$0.011 g/g, respectively.

\section{Investigating the sample by low-background gamma-spectroscopy}
\label{lowbkg}

\subsection{Determination of the $^{238}$U content}

\begin{figure}[htbp]
\begin{center}
\epsfig{figure=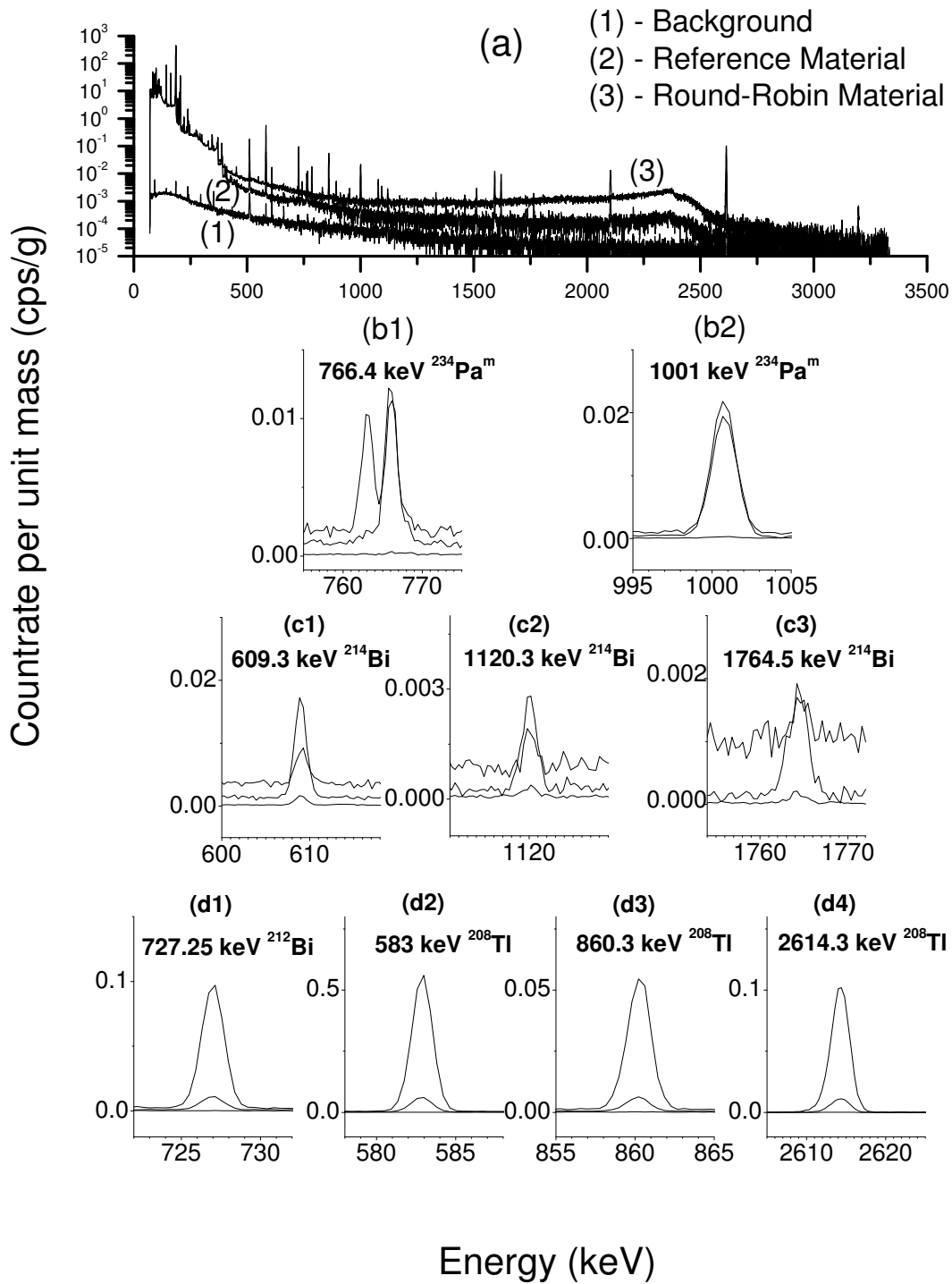, width=\textwidth}
\caption{Gamma-spectrum of the background (1), the reference material (2) and the Round-Robin material (3)}
\label{LowBgSpec}
\end{center}
\end{figure}

The samples of the RFM (1.05 g) and the RRM (0.975 g) were held in a closed cylindrical polyethylene container of 2.9 cm inner diameter as in the measurements described above and assayed at a distance of 6.5 cm from a 150 cm$^3$ coaxial high-purity germanium detector (PGT PIGC 3520 with 34.1 \% relative efficiency) in a low background chamber. Their gamma-spectrum is shown in Fig. \ref{LowBgSpec} together with the background spectrum.

The gamma-peaks at 766.37 and 1001 keV (Fig. \ref{LowBgSpec}b1 and Fig. \ref{LowBgSpec}b2) of $^{234m}$Pa can be observed clearly. Since $^{234m}$Pa is a daughter of $^{238}$U, its 1001 keV gamma-transition was used for estimating the $^{238}$U content. The count rate per unit mass at 1001 keV corresponding to the RRM was compared to the one corresponding to the RFM. Because the sample is ``thin'' and practically transparent for the 1001 keV line, there was no need to correct for self-absorption. Therefore, the count rates per unit mass  at 1001 keV $K_0(1001keV)$ and $K'_0(1001keV)$ were evaluated by simply dividing the measured count rate by the total mass of the sample. Inserting the received values (see Table \ref{tab:U-content}) into a formula analogous to formula (\ref{M}), the $^{238}$U content of 1 g sample was estimated to be 0.065$\pm$0.002 g/g.

\subsection{Age determination}\label{age}

As reported in Refs. \cite{Tam,TamZsigrai}, the activity ratio $^{214}$Bi/$^{234}$U can be used as a chronometer for uranium age dating. The gamma-transitions at 609.3, 1120.3 and 1764.5 keV were identified for $^{214}$Bi (Fig. \ref{LowBgSpec}c1-c3). The activity of $^{214}$Bi in 1g of the sample is calculated by the formula:
\begin{equation}
A={K\over{B\epsilon}} \label{Biactivity}
\end{equation}
where $K$ is the count rate of 1g of the sample, $B$ is the emission probability and $\epsilon$ is the detector efficiency measured by point-like standard sources. Because of using a thin sample, the count rate does not need to be corrected for self-absorption in the sample.

The activity of $^{214}$Bi in the RRM was calculated in Ref. \cite {Tam} and it is summarized in Table \ref{tab:activities}. $^{214}$Bi is a daughter of $^{234}$U, which decays through $^{230}$Th to $^{226}$Ra, which, in turn, decays to $^{214}$Bi through three short-lived nuclides. If the activity of $^{214}$Bi, $A$, and the activity of $^{234}$U, $A_0$, are known, the age $T$ of the sample can be derived from the following formula \cite{Tam,TamZsigrai}:
\begin{equation}
{A\over {A_0}} = {1\over 2}\lambda_{Th}\lambda_{Ra} T^2 \ ,\label{quadraticage}
\end{equation}
where $\lambda_{Th} =0.288\times 10^{-12} s^{-1}$ and $\lambda_{Ra} = 13.86\times 10^{-12} s^{-1}$ are the decay constants of $^{230}$Th and $^{226}$Ra \cite{TORI}, respectively. The age of the RRM was estimated in this way to be $23\pm 3$ years. The age of the RFM was also calculated and it was found to be $42\pm 3$ years, which was the expected value.

\subsection{Determining the activity of $^{232}$U and $^{232}$Th}

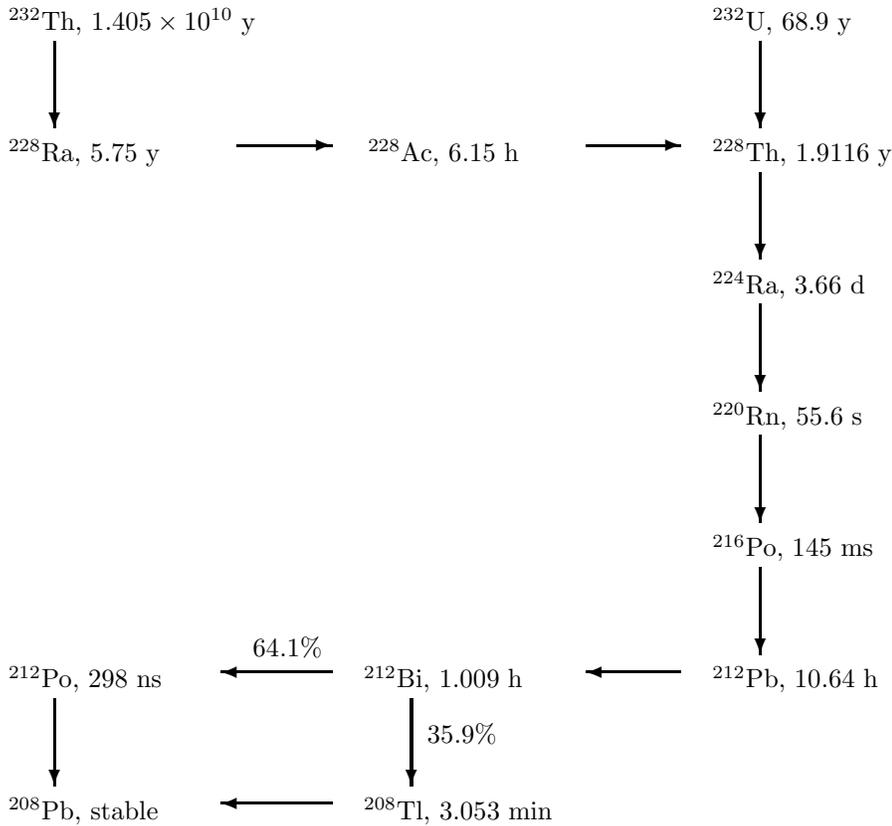
\begin{figure}[htbp]
\begin{center}
\setlength{\unitlength}{0.06pt}
\begin{picture}(5800,5400)(0,-5200)
\thicklines
\put(60,0){$^{232}$Th, $1.405\times 10^{10}$ y}
\put(350,-70){\vector( 0,-1){550}}
\put(60,-830){$^{228}$Ra, 5.75 y}
\put(1500,-730){\vector( 1, 0){600}}
\put(2326,-830){$^{228}$Ac, 6.15 h}
\put(3700,-730){\vector( 1, 0){600}}
\put(4500,0){$^{232}$U, 68.9 y}
\put(4800,-70){\vector( 0,-1){550}}
\put(4500,-830){$^{228}$Th, 1.9116 y}
\put(4800,-900){\vector( 0,-1){550}}
\put(4500,-1660){$^{224}$Ra, 3.66 d}
\put(4800,-1730){\vector( 0,-1){550}}
\put(4500,-2490){$^{220}$Rn, 55.6 s}
\put(4800,-2560){\vector( 0,-1){550}}
\put(4500,-3320){$^{216}$Po, 145 ms}
\put(4800,-3390){\vector( 0,-1){550}}
\put(4500,-4150){$^{212}$Pb, 10.64 h}
\put(4300,-4050){\vector( -1,0){600}}
\put(2300,-4150){$^{212}$Bi, 1.009 h}
\put(2100,-4050){\vector( -1,0){700}}
\put(1600,-3950){64.1\%}
\put(2600,-4220){\vector( 0,-1){550}}
\put(2700,-4500){35.9\%}
\put(60,-4150){$^{212}$Po, 298 ns}
\put(2300,-4980){$^{208}$Tl, 3.053 min}
\put(350,-4220){\vector( 0,-1){550}}
\put(2100,-4880){\vector( -1,0){700}}
\put(60,-4980){$^{208}$Pb, stable}
\end{picture}
\caption{Decay scheme of $^{232}$Th and $^{232}$U.}
\label{Fig:decay}
\end{center}
\end{figure}

During the operation of a nuclear reactor the isotope $^{232}$U is formed. Therefore, if the presence of $^{232}$U  is detected in the investigated material, it is probably a re-processed material produced from spent nuclear fuel. The activity of $^{232}$U  is determined from the activity of its gamma-emitting daughters, $^{212}$Bi and $^{208}$Tl. However, since these isotopes are also present in the decay chain of $^{232}$Th (Fig. \ref{Fig:decay}), determining the activity of $^{232}$U also involves the determination of the activity of $^{232}$Th.

Observing the decay scheme of $^{232}$U and $^{232}$Th (Fig. \ref{Fig:decay}), and the gamma energies emitted by their daughters, it can be seen that the activity of $^{232}$Th can be calculated from the activity of $^{228}$Ac, which, in turn, can be determined from the gamma peaks of $^{228}$Ac at 911.204 and 968.971 keV. Using the law of radioactive decay one obtains
\begin{equation}
A_{Ac228}=A_{Ra228}=A_{Th232}(1-e^{-\lambda_{Ra228}t}) \label{Ac-activity}
\end{equation}
where $A_{Ac228}$, $A_{Ra228}$ and $A_{Th232}$  denote the corresponding activities, $\lambda_{Ra228}$ is the decay constant of $^{228}$Ra, while $t$ is the age of the sample.

The activity of $^{232}$U can be calculated from the activity of $^{212}$Bi and $^{208}$Tl, which can be determined from the count rates at their gamma peaks. The presence of these isotopes in the sample, however, may due both to the decay of $^{232}$Th and to the decay of $^{232}$U. Therefore, the activity of  $^{212}$Bi and of $^{208}$Tl can be calculated as the sum of two terms, one of them accounting for the buildup from $^{232}$Th and the other term accounting for the buildup from $^{232}$U:

\begin{eqnarray}
A_{Bi212}={A_{Tl208}\over f}&=&A_{Th232}\Biggl[1-{{\lambda_{Th228}e^{-\lambda_{Ra228}t}-
\lambda_{Ra228}e^{-\lambda_{Th228}t}}\over{\lambda_{Ra228}-\lambda_{Th228}}}\Biggr]+ \nonumber\\
&+&A_{U232}\Biggl[{{1-e^{(\lambda_{U232}-\lambda_{Th228})t}}\over{1-\lambda_{U232}/\lambda_{Th228}}}\Biggr]\ , \label{BiTl-activity}
\end{eqnarray}
where $f=35.9\% $ is the decay branching probability of the $^{212}$Bi $\rightarrow$ $^{208}$Tl decay (see Fig. \ref{Fig:decay}). Using the measured activities of $^{228}$Ac, $^{212}$Bi and $^{208}$Tl, the measured age of the sample and the known half-lifes, the activities of $^{232}$U and $^{232}$Th are obtained by solving equations (\ref{Ac-activity}) and (\ref{BiTl-activity}) for $A_{U232}$ and $A_{Th232}$.

The activities of $^{212}$Bi and $^{208}$Tl were determined from the measured count rates using the known efficiency of the detector (in the same way as the activity of $^{214}$Bi in section \ref{age}) and they are presented in Table \ref{tab:activities}. The activity of $^{228}$Ac can be estimated from its peaks at 911.204 and 968.971 keV. After subtraction of the background, however, the net count rates of the 911.204 and 968.971 keV peaks turned out to be zero within the measurement error, both in the spectrum of the RRM and of the RFM. Therefore, the upper limits of these count rates may be taken to be equal to their statistical errors. This leads to an upper limit for the value of the $^{232}$Th activity of about 0.37 Bq/g in the RRM and 0.19 Bq/g in the RFM. Inserting these values into the first term in formula (\ref{BiTl-activity}), one obtains 0.40 Bq/g and 0.19 Bq/g as the contribution of the $^{232}$Th decay to the activity of $^{212}$Bi in the RRM and in the RFM, respectively. Since these values are much smaller than the uncertainty of the 
measured $^{212}$Bi and $^{208}$Tl activities, the first term in equation (\ref{BiTl-activity}) can be safely neglected. Therefore, inserting the weighted average of the measured values of $A_{Bi212}$ and $A_{Tl208}/f$ into the left-hand side of formula (\ref{BiTl-activity}) one obtains the $^{232}$U activity as $419\pm 8$ Bq/g for the RRM and $42\pm 1$ Bq/g for the RFM.

\section{Determination of $^{236}$U content and $^{241}$Am activity by a medium area planar HPGe detector}
\label{U236}

\begin{figure}[htbp]
\begin{center}
\epsfig{figure=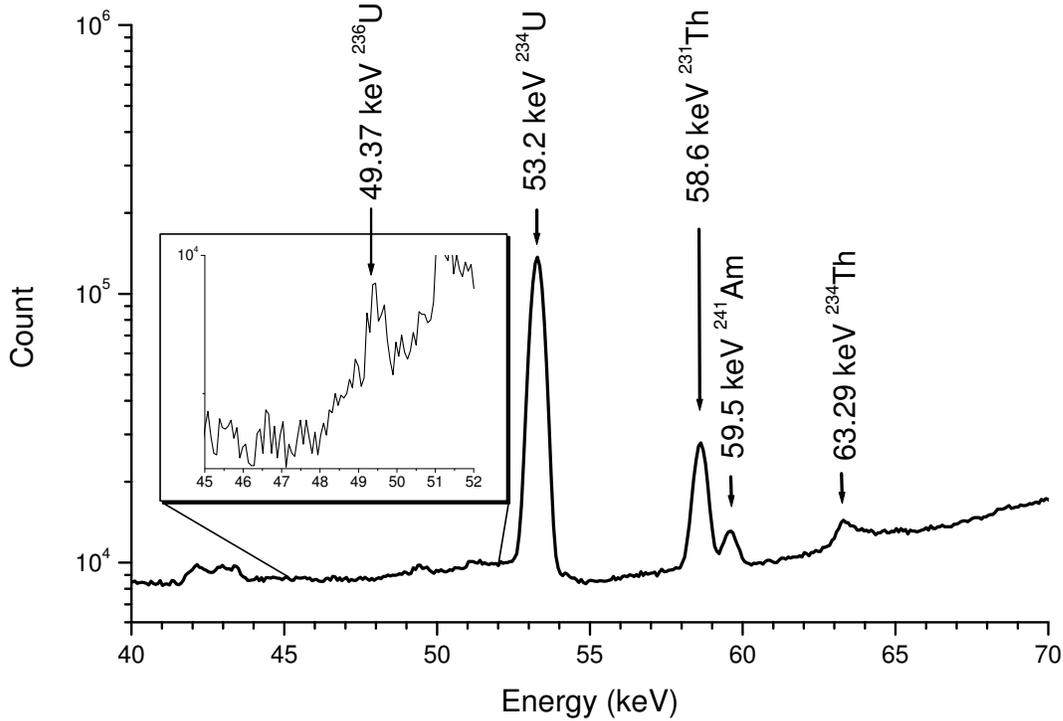, width=\textwidth}
\caption{Gamma-spectrum of the Round-Robin material taken by a medium-area planar HPGe detector.}
\label{Fig:u236}
\end{center}
\end{figure}

The spectrum of the RRM was also taken by a medium area planar HPGe detector, ORTEC SGD-36550-S, with a crystal diameter of 36.7 mm. The 49.37 keV peak observed in the spectrum was identified for $^{236}$U decay (Fig. \ref{Fig:u236}). Although this peak is very close to the 49.550 keV peak of $^{238}$U, in the present case this does not disturb the evaluation of the 49.37 keV peak. Namely, in the observed sample the isotopic ratio of $^{238}$U is so low that the intensity of its 49.550 keV peak is less than the background level, i.e. it is below the detection limit. Therefore, the $^{238}$U peak at 49.550 keV can be neglected when evaluating the one at 49.37 keV.

The ratio of $^{236}$U and $^{235}$U contents was evaluated by comparing the yield of the $^{236}$U peak at 49.37 keV to the yield curve obtained from the yields of the 58.6, 84.2 and 90 keV peaks of $^{235}$U and it was found to be $(0.67\pm 0.10)\times10^{-2}$. Then the $^{236}$U content was calculated to be $(0.51\pm 0.08)\times 10^{-2}$ g/g for the RRM.

Similarly, the 59.6 keV line was identified for $^{241}$Am and its activity was calculated by peak ratio technique as above and found to be $125\pm15$ Bq/g for the RRM.

In the spectra of the RFM the peaks of $^{236}$U and $^{241}$Am could not be identified. More precisely, the upper limit of the count rates at the peaks of $^{236}$U and $^{241}$Am may be estimated as the statistical error of the count rate in the areas of the spectrum where these peaks should be (at 49.37 keV and at 59.6 keV). In this way, for the RFM the upper limit of the $^{236}$U content was estimated to be about  0.0008 g/g, while the $^{241}$Am activity in the RFM was estimated to be not more than 15 Bq/g.

\section{Discussion}
\label{discussion}

In the present paper independent NDA methods for HEU characterization are described. These methods were developed within the two-months time frame of the Round Robin Exercise organized in 2001 by the International Technical Working Group for Combating Illicit Trafficking of Nuclear Materials \cite{Dudderetal,Dudderetal-Karlsruhe}. In fact, the results for the $^{235}$U content and total uranium content were included already in the 24-hours report. The results of the detailed calculations of the $^{234}$U, $^{235}$U, $^{236}$U, $^{238}$U and $^{232}$U contents were used for preparing the final, two-months report, together with the results of mass spectrometric measurements. The age of the Round-Robin material determined by the method described here was also available before the end of the two-months time frame of the exercise. The results of the presented research are compared to the results obtained by mass spectrometry in Table \ref{tab:RR-results}.

Age dating is one of the most interesting results of the presented research. In the Round Robin Exercise in 2001 the age of HEU was estimated by gamma spectrometry for the first time. This method does not require the use of calibration standards and its uncertainty is comparable to that of mass-spectrometry (see Table \ref{tab:RR-results}). Age dating requires measuring the content of $^{234}$U and $^{214}$Bi as precisely as possible. Unlike commercially available software \cite{U235,MGAU,MGA++} (see also \cite{BerlizovTryshyn}), the procedure described in this work can provide a quantitative estimation of the $^{234}$U content which is sufficiently precise for age dating.

The estimation of total uranium content is an important result of this analysis. The obtained value of $0.835\pm0.015$ g/g is close to the uranium content of U$_3$O$_8$ (0.846 g/g). But if we note that the amounts of the three uranium isotopes ($^{234}$U, $^{235}$U and $^{238}$U) in the RRM are significantly smaller than in the RFM (Table \ref{tab:U-content} or Fig. \ref{kev121and185} and Fig. \ref{LowBgSpec} b1-b2), which is known to be U$_3$O$_8$, it follows that the matrix of the RRM is different from the matrix of the RFM and it cannot be identified as U$_3$O$_8$ .

The ability to determine a small $^{238}$U amount in a low background chamber makes it possible to precisely determine the isotopic composition of a material even with nearly 100\% of $^{235}$U. 

Estimating the $^{232}$U and $^{236}$U content provides important information about unknown nuclear material. Because the nuclides of $^{232}$U and $^{236}$U are produced during the operation of a nuclear reactor, their existence in the investigated sample shows that the HEU material was produced from reprocessed spent nuclear fuel. A good energy resolution planar HP-Germanium (small or medium area) should be used to observe the 49.37 keV peak of $^{236}$U  for improving the accuracy of the result.

The accuracy of determining the age of HEU and the ratio of major uranium isotopes ($^{234}$U, $^{235}$U and $^{238}$U) by gamma spectroscopy is comparable to the accuracy of mass spectroscopy. Furthermore, gamma-spectroscopy can also provide information on the total uranium content of the sample, which cannot be determined by mas spectrometry. The isotopic ratio of $^{236}$U, however, which is easily measured by mass spectrometry, is difficult to resolve by gamma spectroscopy, and below a certain limit (around 0.1\%) its gamma-spectroscopic determination becomes impossible. On the other hand, only gamma spectroscopy can be used to measure small activities of $^{232}$U. Since $^{236}$U and $^{232}$U are produced during the operation of a nuclear reactor, identifying the presence of either of them is sufficient to conclude that the sample was produced from reprocessed spent nuclear fuel.  

Ideally, for the full attribution of nuclear materials of unknown origin a balanced interaction between non-destructive and destructive techniques should be achieved. It has been shown here that high-resolution gamma-spectroscopy is a powerful tool of non-destructive analysis for the characterization of highly enriched uranium. Since gamma-spectroscopy is one of the most simple and most cost effective techniques, it might be the first choice for the basic characterization of highly enriched uranium. If a more detailed characterization, including, e.g., full trace element analysis, description of crystal structure, estimation of particle size distribution etc. is needed, other methods of assay, if available, should be called for.

\section*{Acknowledgment}

This research has been partially supported by the Hungarian Atomic Energy Authority under the research and development contract OAH/\'ANI-ABA-75/00. Thanks are due to Laszlo Lakosi for useful suggestions and for carefully checking the manuscript.

\begin{landscape}

\section*{Tables}

\begin{table}[htbp]
\caption{Uranium content and isotopic composition of the Reference Material and of the Round-Robin Material. ($^\#$Activity (Bq/g), $^*$uranium content of U$_3$O$_8$.) }
	\begin{center}
		\begin{tabular}{ccccccc}
			\hline
			Nucleus& \multicolumn{3}{c}{Reference Material}& \multicolumn{3}{c}{Round-Robin Material} \\
			& Isotope ratio (\%) &  Content (g/g)& $K'_0$ (cps/g) & $K_0$ (cps/g) & Content (g/g)& Isotope ratio (\%) \\
			 \hline
			 $^{234}$U & 1.02(4) & 0.0086(3) &  4.9(2) & 4.6(2)& 0.0081(4)& 0.98(7)\\
			 $^{235}$U & 89.5(1.0) & 0.757(12) & 163.0(7) &160.5(7) & 0.745(15)& 89.8(7)\\
			 $^{238}$U & 9.5(2) & 0.080(1) & 0.00351(3)  & 0.00312(3)& 0.0715(35)& 8.6(5)\\
			 \cline{2-7}
			 $^{236}$U & $\le 0.1$ & $\le 0.0008$ &  - & - & 0.0051(8) & 0.6(1)\\
			 $^{232}$U&  -& $5.4(2)\times 10^{-11}$& 42(1)$^\#$& 419(8)$^\#$& $5.4(1)\times 10^{-10}$& - \\
			 \hline
			 Total & & 0.846$^*$ & & & 0.830(20) \\
			\hline
		\end{tabular}
	\end{center}
	\label{tab:U-content}
\end{table}

\pagebreak

\begin{table}[htbp]
\caption{Activity of $^{234}$U, $^{214}$Bi, $^{212}$Bi and $^{208}$Tl. Emission probabilities were taken from Ref. \cite{TORI}, except for 120.9 keV of $^{234}$U, which was taken from Ref. \cite{IAEAdata}.}
	\begin{center}
		\begin{tabular}{cccccccc}
			\hline
			& & & & \multicolumn{2}{c}{Round-Robin Material}& \multicolumn{2}{c}{Reference Material}\\
			Nucleus& \parbox{1.5cm}{\center Energy\\ (keV)}& \parbox{2cm}{\center Emission \\probability\\ (\%)}&\parbox{2cm}{\center Detector \\ efficiency (\%)}&\parbox{2.5cm}{\center Count rate \\per unit mass \\($\times$100 cps/g)}& \parbox{2.5cm}{\center Activity per unit mass (Bq/g)}&\parbox{2.5cm}{\center Count rate \\per unit mass \\($\times$100 cps/g)}& \parbox{2.5cm}{\center Activity per unit mass (Bq/g)}\\
			\hline
			 $^{234}$U & 120.9 &0.0342  &0.730(15)&460(15)  &1.84(6)$\times 10^6$ \cite{Tam}& 490(15)& $1.91(6)\times 10^6$ \cite{Tam}\\
			 \hline
			 $^{214}$Bi &   609.3, 1120.3, 1764.5 & & & Average= & 2.00(25) \cite{Tam}& Average=& 7.06(15)\cite{Tam}\\
			 \hline
			 $^{212}$Bi &  727.3&6.58  &0.483(1) &13.6(1)&429(15) &1.38(5)& 43(3)\\
			 \hline
			 $^{208}$Tl & 583.0 &84.5  &0.58(1)&76.2(2) &156(4)&7.72(15)& 16(1)\\
			 &860.3&12.14 &0.45(1)&8.37(8)&153(6)&0.76(5) &14(1)\\
			 & 2614.3& 99 & 0.15(2) & 25.5(1)&172(25)&2.46(5)& 17(3)\\
			 & & & & Average=& 155(4)& Average=& 15(1)\\
			\hline
		\end{tabular}
	\end{center}
	\label{tab:activities}
\end{table}

\pagebreak

\begin{table}[htbp]
\caption{Isotopic composition, $^{232}$U content and the age (in 2001) of the Round-Robin Material, measured by different laboratories using mass-spectrometry (MS)  and gamma-spectrometry (GS) \cite{Dudderetal,Dudderetal-Karlsruhe}.}
	\begin{center}
		\begin{tabular}{cccccccc}
		\hline
		Laboratory nickname & $^{234}$U & $^{235}$U & $^{236}$U & $^{238}$U & $^{232}$U (g/g) & Age (years) & Method \\
		\hline
		Azores &	0.97	&	89.99	&	0.68	&	8.37	&	&		&	MS,GS\\
		 &		&		&		&		&	&		22.2-22.6	&	MS\\
		Barbados &		& $85.6\pm 3.8$	&		&	&	&	&		GS\\
		Borneo	&	0.85 $\pm$  0.15	&	86.7  $\pm$ 1.5	&	0.57 $\pm$  0.08	&	11.9  $\pm$ 0.9&	&	&	MS\\
		Chatham	&	0.960 $\pm$   0.001&	89.94  $\pm$  0.06&	0.0643 $\pm$ 0.003&	8.462  $\pm$  0.006	&	&	&			MS\\
		Galapagos	&	0.96	&	89.89	&	0.68	&	8.47	&	&	&			MS\\
		Mindanao&	0.96 $\pm$ 0.40	&	89.9 $\pm$ 0.11&	0.678 $\pm$  0.23	&	8.443 $\pm$ 1.29&	&	22.4 $\pm$ 1.2&	MS\\
		Tobago	&	1.05$\pm$0.07	&	89.37$\pm$1.8	&	0.69$\pm$0.05	&	8.88$\pm$0.2&	&	&	MS\\
		Tonga	&	0.967$\pm$ 0.001	&	89.99 $\pm$ 0.02	&	0.679 $\pm$0.001	&	8.362 $\pm$ 0.005&	&	&	MS\\
Trinidad	&	0.955$\pm$0.075	&	90.01$\pm$0.35	&	0.673$\pm$0.030	&	8.365$\pm$0.033	&	&	23.5$\pm$ 0.5&	MS\\
\hline
Present work	&	0.98$\pm$0.07	&	89.8$\pm$0.7	&	0.6$\pm$0.1	&	8.6$\pm$0.5	&	$(5.4\pm 0.1)\times 10^{-10}$	&23$\pm$3&		GS\\

		\hline
		\end{tabular}
	\end{center}
	\label{tab:RR-results}
\end{table}

\end{landscape}

\end{document}